\documentclass[preprint,12pt]{elsarticle}
\usepackage{amssymb}
\journal{Physica E}

\begin{document}

\begin{frontmatter}

\title{Quantum plateaus in dynamical Hall conductivity}

\author{ Zhyrair Gevorkian$^{1,2,*}$, Vadim Farztdinov$^{3}$, Yurii Lozovik$^{3,4}$ }
\address{$^{1}$ Yerevan Physics Institute,Alikhanian Brothers St. 2,0036 Yerevan, Armenia.\\
$^2$Institute of Radiophysics and Electronics,Ashtarak-2,0203,Armenia\\
$^{3}$ Institute of Spectroscopy, Academy of Sciences of Russia,
142092 Troitsk, Moscow Region. Russian Federation\\
$^{4}$  Moscow Institute of Physics and Technology, Moscow Region, Dolgoprudny, Russia }

\begin{abstract}
Dynamical Hall conductivity $\sigma_H(\omega)$ of a $2D$ electron gas with impurities  in the perpendicular magnetic field is analyzed. Plateau-like behavior at low frequencies as well as at high frequencies provided the complete filling of Landau levels is predicted. The broadening of a Landau level separates two frequency regions with different behaviour.  Imaginary part of dynamical Hall conductivity reveals oscillations in the localized states region. Comparison with the experiment is carried out.
\end{abstract}

\end{frontmatter}

\section{Introduction}
The discovery of the Quantum Hall Effect \cite{klitzing} triggered theoretical \cite{LFG84,AL85,ALC85} and experimental \cite{kuchar86,galchen87} studies dedicated to investigation of an electric field frequency influence on quantization of the  Hall conductivity of $2D$ electron gas in perpendicular magnetic field.
 The main question here was to discover what is happening with the static Hall conductivity plateau when increasing the external electric field frequency. In early experiments \cite{kuchar86,galchen87} the survival of plateau up to several ten GHz frequencies was found. This result was corroborated in a recent experiment  \cite{dziom18}. Besides, the existence of plateau even at high THz frequencies was reported \cite{ikebe10,stier15,failla16} recently. Trying to explain the above mentioned experimental features we consider in detail this problem theoretically by using a simple model of $2D$ electron gas with a single $\delta$-shaped impurity potential in a perpendicular magnetic field \cite{bem78, prange81, LFG84}.
 Note that in most experiments the strength of magnetic field varies in the 1T $-$ 6T range. The corresponding magnetic length $r_H=\sqrt{c\hbar/eH}$
is in the range 26nm $-$ 10nm and is essentially larger than a radius of impurity potential, the latter as a consequence can be considered as infinitesimally small. Therefore the model with $\delta$-shaped impurity potential seems to be more adequate as compared to other theoretical models \cite{AL85,ALC85,mor09,mor10} which potentially are more relevant for magnetic field of larger strength. Below we show  that within this theoretical model the plateau-like behaviour can be understood both in low microwave and in large terahertz frequency ranges.
\section{Model Description}
Let us consider 2D electron in a perpendicular magnetic field  with $\delta$-shaped impurity potential \cite{bem78,prange81}. The Hamiltonian of the system has the form
\begin{equation}
\hat{H}=\hat{H_0}+2\pi\lambda\delta(\vec r),\quad \hat{H_0}=-\frac{1}{2}\frac{\partial^2}{\partial y^2}+\frac{1}{2}(p_x+y)^2
\label{ham}
\end{equation}
For convenience here we use the following units $\hbar=m=e=c=H=1$ and the vector potential reads $\vec A\equiv (-y,0,0)$ \cite{prange81}. Note that in the above mentioned paper only the static case $\omega=0$ was considered. In these units magnetic length and cyclotron frequency $r_H\equiv 1$, $\omega_c\equiv 1$. 
The spectrum of Hamiltonian consists of two types of states: delocalized states which are not split off from a Landau level, and localized at the impurities states which are split off from Landau level, see also \cite{bem78,prange81,LFG84}. 
The number of localized states equals to the number of impurities ($N_{imp}$) provided that their number is significantly lower than a Landau level degeneracy  $N_{imp}\ll N_0\equiv S/2\pi$, where $S$ is the area of the system. We will consider the case of a single impurity and then qualitatively generalize the results to the case of many impurities. 

The wave function of delocalized state has the form \cite{bem78,prange81}
\begin{equation}
\psi_{\alpha n}(\vec r)=\sum_{p}c_{pn}^\alpha\psi_{p n}(\vec r),\quad E_{\alpha}=n,\quad \sum_{p}c_{pn}^\alpha\psi_{p n}(0)=0,\quad \alpha=1,...N_0-1
\label{unsplit}
\end{equation}
where $\psi_{np}$ are the eigenstates of $H_0$, \cite{landau}
\begin{equation}\label{landau}
\psi_{np}(\vec r)=\frac{\exp(ipx)H_n(y+p)e^{-(y+p)^2/2}}{\pi^{1/4}\sqrt{L2^n n!}}
  \end{equation}
Here $H_n$ are Hermite polynomials and $L$ is the system size in $x$ direction. 
The last sum in Eq.(\ref{unsplit}) reflects condition imposed on delocalized state wave function that it turns to zero at the origin and therefore does not feel the impurity potential and stays delocalized.  
An explicit expression for expansion coefficients $c_{pn}^\alpha$ will not be required. Instead, further below we will show that these coefficients obey certain sum rules, which allow to get simple analytical expressions for the conductivity. 

The wave function and energy of a localized state  are determined as \cite{prange81}
\begin{eqnarray}\label{loc1}
  \psi_R(\vec r)=\sum_{np}c_{np}^R\psi_{np}(\vec r),\quad c_{np}^R=A_R\frac{\psi_{np}^*(0)}{E_R-n}\\\nonumber  A_R^{-2}=\frac{1}{2\pi}\sum_{n=0}^{\infty}\frac{1}{(E_R-n)^2}=\frac{1}{2\pi}\psi_1(-E_R),\quad \lambda\sum_{n=0}^{N_0-1}\frac{1}{E_R-n}=1
  \end{eqnarray}
where $\psi_1(z)$ is the trigamma function \cite{Abst}. 
The formally divergent sum in dispersion equation Eq.(\ref{loc1}) is truncated by the number of states in a system of finite size \cite{prange81}. Its solution shows that from each degenerate level one state is split off, the direction follows the sign of $\lambda$. The total number of new states is $N_0$. The energy of the split off state can be written as $E_R = n + \varepsilon_n$. Using the properties of trigamma function one can show that 
\begin{equation}\label{Asquared}
  A_R^{-2} = \frac{1}{2\pi}\left[\frac{1}{\varepsilon_n^2}+\frac{\pi^2}{6}C_n + O(\varepsilon_n)\right]
\end{equation}
where $C_n$ is slowly varying coefficient $1\le C_n \le 2$, and $C_0=1, C_\infty=2$. 

Substituting in the first expression of Eq.(\ref{loc1}) the coefficients  $c_{np}^R$ by their values (the second expression in Eq.(\ref{loc1})) and taking the sum over $p$ by integration $\sum_{p}\rightarrow \frac{L}{2\pi}\int dp$, one obtains
\begin{equation}\label{moment}
  \psi_R(\vec r)=\frac{A_R}{\sqrt{2\pi}}\sum_{n}\frac{\psi_{nm=0}(\vec r)}{E_R-n},\quad \psi_{nm=0}(\vec r)=\sqrt{2\pi}\sum_{p}\psi_{np}(0)\psi_{np}(\vec r)
\end{equation}
where $\psi_{nm}$ are eigenstates of Hamiltonian $H_0$ in the angular momentum representation \cite{landau}
\begin{eqnarray}\label{momrep}
  &\psi_{nm}(\vec r)=&\left[\frac{n_r!}{2\pi(n_r+|m|)!}\right]^{1/2}\exp(-i m\Phi-\frac{x^2+y^2+2ixy}{4})(\frac{x^2+y^2}{2})^{|m|/2}L_{n_r}^{|m|}(\frac{x^2+y^2}{2}) \nonumber\\
  & n_r=&n+\frac{m-|m|}{2},\quad m\geq -n
\end{eqnarray}
Here $L_{n}^{|m|}$ are associated Laguerre polynomials. The last expression in Eq.(\ref{moment}) reflects the fact that both sets of functions $\psi_{nm}$ and $\psi_{np}$ are eigenstates of the same Hamiltonian $H_0$ and each function of one system can be represented as superposition of eigenfunctions of the other. Note also that eigenfunctions $\psi_{nm}$ (Eq.(\ref{momrep})) differ from those in Ref.\cite{landau} by a phase factor $\exp(-ixy/2)$, which takes into account the difference between symmetric gauge and used here Landau gauge. 

As it follows from Eq.(\ref{moment}) only the states with
$m= 0$ split off from Landau level and contribute to the $\psi_R(\vec r)$. These wave functions according to Eq.(\ref{momrep}) are localized around the point $r= 0$.
 All states with $m\neq 0$ remain unsplit and are eigenstates of the Hamiltonian $H$ as well. Therefore the states $|\alpha n\rangle$ are linear combinations of $|n m\neq 0\rangle$
\begin{equation}\label{alpham}
  \psi_{\alpha n}(\vec r)=\sum_{m\neq 0}c_{mn}^{\alpha}\psi_{n m\neq 0}(\vec r), \quad \sum_{\alpha}c_{mn}^{*\alpha}c_{m^{\prime}n}^{\alpha}=\delta_{m m^{\prime}}
\end{equation}
\section{Sum rules}
In this paragraph we will use the completeness  and orthonormality relations to derive sum rules which will be instrumental for the calculation of conductivity further below. The states $|\alpha n>$ and $|n m=0>$ are the eigenstates of the Hamiltonian $H_0$ and form a complete set
\begin{equation}
\sum_{\alpha n}\psi_{\alpha n}^*(\vec r)\psi_{\alpha n}(\vec r^{\prime})+\psi_{nm=0}^*(\vec r)\psi_{nm=0}(\vec r^{\prime})=\delta(\vec r-\vec r^{\prime})
\label{compho}
\end{equation}
Inserting expansions Eqs.(\ref{unsplit}) and (\ref{moment}) into Eq.(\ref{compho}), we obtain
\begin{equation}\label{sumrul1}
  \sum_{\alpha}c_{pn}^{*\alpha}c_{p^{\prime}n}^{\alpha}+2\pi \psi_{np}^*(0)\psi_{np^{\prime}}(0)=\delta_{pp^{\prime}}
\end{equation}
Another relation is found from the orthonormality of the states $|\alpha n>$
\begin{equation}\label{orthonorm}
  \sum_{p}c_{pn}^{*\alpha}c_{pn}^{\alpha^{\prime}}=\delta_{\alpha \alpha^{\prime}}
\end{equation}
The  $\psi_{\alpha n}$ and $\psi_R$ are the eigenstates of the Hamiltonian Eq.(\ref{ham}) and also obey the completeness condition
\begin{equation}\label{comph}
  \sum_{\alpha n}\psi_{\alpha n}^*(\vec r)\psi_{\alpha n}(\vec r^{\prime})+\sum_{R}\psi_{R}^*(\vec r)\psi_{R}(\vec r^{\prime})=\delta(\vec r-\vec r^{\prime})
\end{equation}
Substituting Eqs.(\ref{moment},\ref{alpham}) into Eq.(\ref{comph}) and using the completeness of $\psi_{nm}$ states, one obtains
\begin{equation}\label{condr}
  \frac{1}{2\pi}\sum_{R}\frac{A_R^2}{(E_R-n)(E_R-n^{\prime})}=\delta_{nn^{\prime}}
\end{equation}
\section{Conductivity tensor}
The conductivity tensor of $2D$ electron gas is determined through the Kubo formula
\begin{equation}\label{Kubo}
\sigma_{\mu\nu}(\omega)=\frac{i}{S}\sum_{tt^{\prime}}\frac{f_t-f_{t^{\prime}}}{\varepsilon_{t^{\prime}}-\varepsilon_t}
\frac{v_{tt^{\prime}}^{\mu}v_{t^{\prime}t}^{\nu}}{\varepsilon_t-\varepsilon_{t^{\prime}}-\omega+i\delta}
\end{equation}
where $\hat{v}$ is the velocity operator, $t$ are the quantum numbers determining single electron state, $f_t\equiv f(\varepsilon_t)$ is its distribution function and $S$ is the area of the system. In a magnetic field the velocity operators have the form
\begin{equation}\label{velop}
  \hat{v}^x=\hat{p}_x+y,\quad \hat{v}^x=\dot{y}
\end{equation}
It is helpful to use the relation
\begin{equation}\label{help}
  v^y_{tt^{\prime}}=i^{-1}<t|y|t^{\prime}>(\varepsilon_t-\varepsilon_{t^{\prime}}).
\end{equation}
In our case $t \in \{n\alpha, R\}$. Using Eq.(\ref{help}) for the Hall conductivity one has from Eq.(\ref{Kubo})
\begin{eqnarray}
\label{hallyx}
  \sigma_{yx}(\omega) &=& \frac{1}{S}\sum_{nn^{\prime}\alpha\alpha^{\prime}}\frac{(f_{n^{\prime}}-f_n)<n\alpha|\hat{p}_x+y|n^{\prime}\alpha^{\prime}>
  <n^{\prime}\alpha^{\prime}|y|n\alpha>}{n-n^{\prime}-\omega+i\delta} \nonumber \\
   &&+\frac{1}{S}\sum_{n\alpha R}\frac{(f_{R}-f_n)<n\alpha|\hat{p}_x+y|R>
  <R|y|n\alpha>}{n-E_R-\omega+i\delta}+\nonumber
  \\&&+(n\Longleftrightarrow R)
\end{eqnarray}
When deriving Eq.(\ref{hallyx}) we took into account that $<R|y|R^{\prime}>=0$, which follows from the form of localized state wave function, see Eqs.(\ref{moment},\ref{momrep}). Let us now calculate the matrix elements existing in Eq.(\ref{hallyx}). Using Eqs.(\ref{unsplit},\ref{landau}) one has
\begin{equation}\label{matelemuns}
 \sum_{\alpha\alpha^{\prime}} <n\alpha|\hat{p_x}+y|n^{\prime}\alpha^{\prime}><n^{\prime}\alpha^{\prime}|y|n\alpha>=
  (\delta_{n^{\prime}n-1}\frac{n}{2}+\delta_{n^{\prime}n+1}\frac{n+1}{2})\sum_{pp^{\prime\alpha\alpha^{\prime}}}c_{pn}^{*\alpha}
  c_{p^{\prime}n}^{\alpha}c_{pn^{\prime}}^{\alpha^{\prime}}c_{p^{\prime}n^{\prime}}^{*\alpha^{\prime}}
\end{equation}
Employing the sum rules Eq.(\ref{sumrul1}) one finds from Eq.(\ref{matelemuns})
\begin{equation}\label{matel1}
\sum_{\alpha\alpha^{\prime}} <n\alpha|\hat{p_x}+y|n^{\prime}\alpha^{\prime}><n^{\prime}\alpha^{\prime}|y|n\alpha> =(\delta_{n^{\prime}n-1}\frac{n}{2}+\delta_{n^{\prime}n+1}\frac{n+1}{2})(N_0-2)
\end{equation}
We calculate the matrix elements in the second term Eq.(\ref{hallyx}) in a similar manner
\begin{equation}\label{matel2}
 \sum_{\alpha} <n\alpha|\hat{p_x}+y|R><R|y|n\alpha>=\frac{A_R^2}{4\pi}\left[\frac{n+1}{(E_R-n-1)^2(E_R-n)}-\frac{n}{(E_R-n+1)^2(E_R-n)}\right]
\end{equation}
Substituting Eqs.(\ref{matel1},\ref{matel2}) into Eq.(\ref{hallyx}), one finds
\begin{equation}\label{DL}
\sigma_{yx}(\omega)=\sigma_{yx}^D(\omega)+\sigma_{yx}^R(\omega)
  \end{equation}
where $\sigma^{D,R}$ are the contributions of delocalized and  localized states respectively
\begin{eqnarray}
  \sigma_{yx}^D &=&\frac{N_0-2}{2S}\left[\frac{1}{1-\omega+i\delta} +\frac{1}{1+\omega-i\delta}\right]+\nonumber \\
  &&+\frac{1}{2\pi S}\sum_{Rn}f_nA_R^2\left[\frac{n+1}{(E_R-n-1)^2}-\frac{n}{(E_R-n+1)^2}\right]\times\nonumber\\ &&\times \frac{1}{(E_R-n+\omega-i\delta)(E_R-n-\omega+i\delta)} \nonumber\\
\sigma_{yx}^R&=& -\frac{1}{2\pi S}\sum_{Rn}f_RA_R^2\left[\frac{n+1}{(E_R-n-1)^2}-\frac{n}{(E_R-n+1)^2}\right]\times\nonumber\\
   &&\times\frac{1}{(E_R-n+\omega-i\delta)(E_R-n-\omega+i\delta)} \label{hallfinal}
\end{eqnarray}
Note that $\sigma^D$ also includes transitions between delocalized and localized states as well as $\sigma^R$ includes transitions into delocalized states. The Hall conductivity can be split now into real and imaginary parts
 \subsection{Real part of the Hall conductivity}
For the real part of the Hall conductivity one can get the following expressions
\begin{eqnarray}
  Re\sigma_{yx}^D(\omega) &=&\frac{N_0-2}{S(1-\omega^2)}\sum_{n}f_n+\frac{1}{2\pi S}\sum_{Rn}f_nA_R^2\left[\frac{n+1}{(E_R-n-1)^2}- \frac{n}{(E_R-n+1)^2}\right]\times \nonumber\\ &&\times \frac{1}{(E_R-n)^2-\omega^2}\nonumber\\
  Re\sigma_{yx}^R(\omega) &=& -\frac{1}{2\pi S}\sum_{Rn}f_RA_R^2\left[\frac{n+1}{(E_R-n-1)^2}- \frac{n}{(E_R-n+1)^2}\right]\frac{1}{(E_R-n)^2-\omega^2}
  \label{realpart}
\end{eqnarray}
 It follows from Eq.(\ref{realpart})
  that there are two main frequency ranges: 1) a low frequency range $\omega \ll |\varepsilon_n|\equiv|E_R-n|$ , 2) a high frequency range $\omega\gg |\varepsilon_n|$. Remember that $\varepsilon_n$ is the energy of a split off state that describes the broadening of Landau level, and note that we do not consider very high frequencies $\omega\gg \omega_c$ where conductivity is trivially determined by Drude formula.

Let us first consider the low frequency range. Expanding the expressions in Eq.(\ref{realpart}) on $\omega$ and using sum rules Eq.(\ref{condr}) one can find
  \begin{eqnarray}\label{realpartfin}
    Re\sigma_{yx}^D(\omega)&=& \frac{1}{2\pi(1-\omega^2)}\sum_{n}f_n+\frac{\omega^2}{2\pi N_0}\sum_{n}f_n+\frac{\omega^2}{2\pi N_0}\sum_{nR}f_n\frac{A_R^2}{2\pi}\left[\frac{1}{(E_R-n)^4}+\frac{4n+2}{(E_R-n)^3}\right] \nonumber \\
     Re\sigma_{yx}^R(\omega)&=&-\frac{\omega^2}{2\pi N_0}\sum_{R}f_R-\frac{\omega^2}{2\pi N_0}\sum_{nR}f_R\frac{A_R^2}{2\pi}\left[\frac{1}{(E_R-n)^4}+\frac{4n+2}{(E_R-n)^3}\right]
  \end{eqnarray}
When obtaining Eq.(\ref{realpartfin}), using sum rules, we partially take the sums over $R$ in $ Re\sigma_{yx}^D$ and over $n$ in  $Re\sigma_{yx}^R$.
 As can be seen from Eq.(\ref{realpartfin}) in the static case $\omega=0$, $    Re\sigma^R(\omega=0)\equiv 0$ and the filling of only delocalized states is enough for quantization of the Hall conductivity \cite{klitzing}
 \begin{equation}\label{quant}
   Re\sigma_{yx}(0)=\frac{n^*}{2\pi}
 \end{equation}
 where $n^*=\sum_{n}f_n$ is the number of completely filled Landau levels. Although the localized states do not contribute to the static Hall conductivity, the virtual transitions from delocalized to localized states do and are very important to ensure the quantized value of conductivity \cite{klitzing}. In the dynamic case $\omega\neq 0$, provided that Landau levels are completely filled, one has from Eq.(\ref{realpartfin})
 \begin{equation}\label{compfill}
   Re\sigma_{yx}(\omega)=Re\sigma_{yx}^D(\omega)+Re\sigma_{yx}^R(\omega)=\frac{n^*}{2\pi(1-\omega^2)}
 \end{equation}
 Therefore in the dynamic case the ideal value of the Hall conductivity (for impurity-free system) is attained only after complete filling of a Landau level including the localized states. Now let us investigate how the Hall conductivity is changing when filling the localized states takes place. For this goal we will simplify Eq.(\ref{realpartfin}) by keeping only the resonance terms in sums over $nR$
 \begin{eqnarray}\label{res}
    Re\sigma_{yx}^D(\omega)&=& \frac{1}{2\pi(1-\omega^2)}\sum_{n}f_n+\frac{\omega^2}{\pi N_0}\sum_{n}f_n+\frac{\omega^2}{2\pi N_0}\sum_{n}f_n\frac{A_{Rn}^2}{2\pi}\left[\frac{1}{\varepsilon_n^4}+\frac{4n+2}{\varepsilon_n^3}\right] \nonumber \\
     Re\sigma_{yx}^R(\omega)&=&-\frac{\omega^2}{\pi N_0}\sum_{n}f_{Rn}-\frac{\omega^2}{2\pi N_0}\sum_{n}f_{Rn}\frac{A_{Rn}^2}{2\pi}\left[\frac{1}{\varepsilon_n^4}+\frac{4n+2}{\varepsilon_n^3}\right]
 \end{eqnarray}
Previously we have shown (see Eq.(\ref{Asquared})) that $A_{Rn}^2\approx 2\pi\varepsilon_n^2$.
 It follows from Eq.(\ref{res}) that when $\varepsilon_n>0 $ $(\lambda>0)$, the $Re\sigma_{yx}^R(\omega)<0$. 
Therefore the filling of $\varepsilon_n>0 $ localized state results in a decrease of Hall conductivity. 
Note that due to these ($\varepsilon_n>0 $) states the contribution of $Re\sigma_{yx}^D(\omega)$ to the Hall conductivity is larger than in a pure system. 
When $\varepsilon_n<0$ $(\lambda<0)$, the $Re\sigma_{yx}^R(\omega)>0$ if $|\varepsilon_n|>1/(4n+2)$, otherwise the contribution is again negative. Therefore with the filling of $\varepsilon_n<0 $ localized state the Hall conductivity can either increase or decrease depending on the fulfillment of the above mentioned condition.

Now let us consider the high frequency range $\omega\gg |\varepsilon_n|$. Ignoring $\varepsilon_n$ in the denominator  one finds from Eq.(\ref{realpart})
 \begin{equation}\label{high}
   Re\sigma_{yx}^R(\omega)\approx \frac{1}{2\pi N_0}\sum_{n}f_{Rn}\frac{\varepsilon_n^2}{\omega^2}
 \end{equation}
 It follows from Eq.(\ref{high}) that the localized state contribution in the high frequency range is positive independently of the sign of $\varepsilon_n$. Therefore the Hall conductivity will increase, although the increase will be tiny because $\omega\gg \varepsilon_n.$
 \subsection{Imaginary part of the Hall conductivity}
 From Eq.(\ref{hallfinal}) one can find for imaginary part of conductivity the following relationships
 \begin{eqnarray} \label{im}
   Im\sigma_{yx}^D(\omega) &=&\frac{N_0-2}{4N_0}\left[\delta(1+\omega)-\delta(1-\omega)\right]-\frac{1}{8\pi N_0\omega}\sum_{Rn}f_nI_{Rn}\times\nonumber\\
&&\times\left[\delta(E_R-n-\omega)+\delta(E_R-n+\omega)\right] \nonumber\\
   Im\sigma_{yx}^R(\omega)  &=&\frac{1}{8\pi N_0\omega}\sum_{Rn}f_RI_{Rn}\left[\delta(E_R-n-\omega)+\delta(E_R-n+\omega)\right]
 \end{eqnarray}
 where
 \begin{equation}\label{irn}
   I_{Rn}=A_R^2\left[\frac{n+1}{(E_R-n-1)^2}-\frac{n}{(E_R-n+1)^2}\right]
 \end{equation}
 As can be seen from Eq.(\ref{im}) the localized states lead to nonzero imaginary part at frequencies essentially smaller than cyclotron frequency $1$. 
Note that the localized states, split off upward ($\varepsilon_n>0 $) and split off downward ($\varepsilon_n<0 $), provide different contribution to the  imaginary part of the Hall conductivity. 
To prove this let us consider the contribution of localized states into the imaginary part of the Hall conductivity in the resonance approximation
 \begin{equation}\label{imres}
   Im\sigma_{yx}^R(\omega)=\frac{1}{8\pi N_0\omega}\sum_{n}f_{Rn}I_{Rn}(\varepsilon_n)\left[\delta(\varepsilon_n-\omega)+\delta(\varepsilon_n+\omega)\right]
 \end{equation}
 The $\varepsilon_n>0$ states (split off upward) make a contribution into the first $\delta$ function and the $\varepsilon_n<0$ states (split off downward) make a contribution into the second $\delta$ function. 
The signs of these contributions depend on the sign of $I_{Rn}(\pm\omega)$. Using Eq.(\ref{irn}) it is easy to find out that $I_{Rn}(\omega)>0$ and $I_{Rn}(-\omega)<0$ provided that $n\neq 0$ and $\omega<1$. This means that when filling the localized states the imaginary part of the Hall conductivity will oscillate depending on whether the localized state is split off upward or downward. Similar behavior was observed in a recent experiment \cite{dziom18}.

\section{Comparison with experiment and discussion}
 To get a better comparison with experiments let us now qualitatively generalize the results of a single impurity case to the case of many impurities $1\ll N_{imp}\ll N_0$. 
Above we considered low ($\omega\ll \varepsilon_n$) and high ($\omega\gg \varepsilon_n$) frequency ranges. When comparing with experiment, one should keep in mind that when $\omega \lesssim \Delta_n$, where $\Delta_n$ is the broadening of a Landau level, both cases can realize simultaneously, with $\omega\ll \varepsilon_n$ for some part of the impurities and $\omega\gg \varepsilon_n$ for the other. Exact response will be dependent on distribution of impurities (defining the shape of the Landau level broadening). The other point that has to be kept in mind is that changing of Landau level filling in most of real life experiments is achieved through changing the strength of the magnetic field. This means that localized states splitting will be magnetic field dependent and can change from weak splitting at a low number of filled Landau levels to strong splitting at a high one.   
 
In the low frequency range $\omega\ll \Delta_n$, where $\Delta_n$ is the broadening of a Landau level ($\Delta_n\equiv \varepsilon_n$ in case of single impurity   with $\delta$-shaped potential), the real part of dynamical Hall conductivity decreases when electrons fill split off upward localized states from $n$th Landau level. 
When a Landau level is completely filled, the $Re\sigma_{yx}(\omega)$ is equal to its value in an ideal impurity-free system. 
This is a generalization of the quantum Hall effect to dynamic case. Close to a complete filling of Landau level the correction to the quantized value is of the order of $\omega^2/\omega_c^2$  as it follows from Eq.(\ref{compfill}). 
Therefore for the low frequencies $\omega\ll \Delta\ll \omega_c$ the filling of these localized states will not change the Hall conductivity and a plateau-like behaviour will be retained, see Fig.1.
For the experimental values $\omega\sim70GHz$,\quad  $\omega_c\sim 2THz$,\quad $\omega^2/\omega_c^2\sim 10^{-3}$ \cite{kuchar86,galchen87,dziom18}.

\begin{figure}
 \begin{center}
\includegraphics[width=8.0cm]{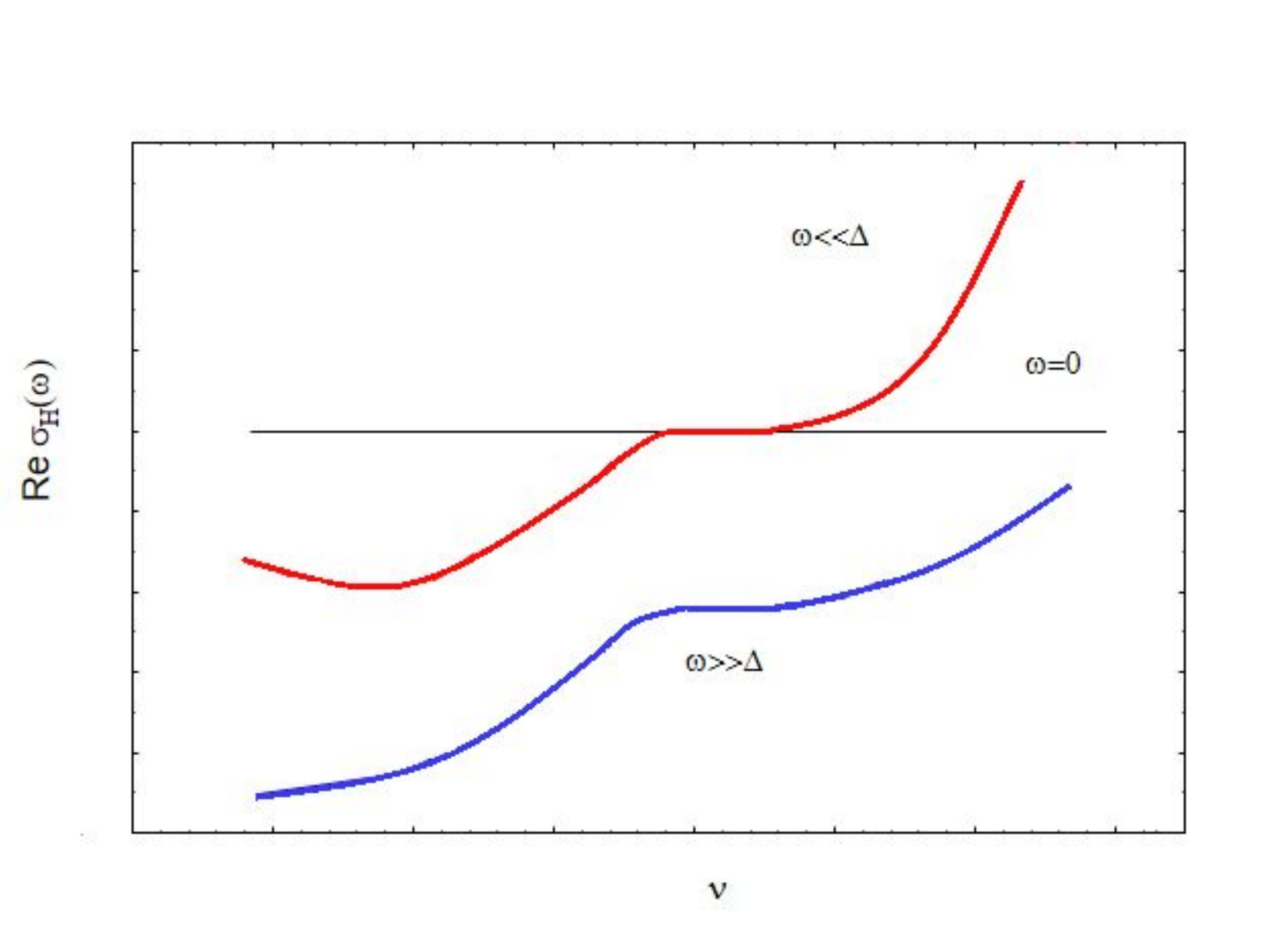}

\caption{Qualitative dependence of real part of dynamical Hall conductivity $Re\sigma_{H}(\omega)$ on Landau level filling factor $\nu$ in the localized states region for three frequency ranges: $\omega \ll \Delta$ -- red line, $\omega = 0$ -- black line, $\omega \gg \Delta$ -- blue line.}
\label{fig.1}
\end{center}
\end{figure}

 When electrons fill the localized $\varepsilon_n<0$ states, which are split off downward from $n+1$-th Landau level, the $Re\sigma_{yx}^R(\omega)$ will decrease or increase depending on the fulfillment of the condition $|\varepsilon_n|>\omega_c/(4n+2)$. 
When this condition is  fulfilled, one will have an increase of the Hall conductivity with filling these states.  A similar behavior was found for slowly varying in space impurity potential \cite{AL85, ALC85}.

For the imaginary part of dynamical Hall conductivity $Im\sigma_{xy}(\omega)$ in the localized state region between two Landau levels the oscillations are predicted due to the different contribution from localized states, split off upward or downward. 
These oscillations were observed in the experiment \cite{dziom18}. As was explained above the Landau level broadening determines the threshold frequency that divides low frequency and high frequency ranges. If one accepts that Landau level broadening in GaAs/AlGaAs heterojunction is caused by short-range ionized impurities (see for example, \cite{broad}), then the Landau level broadening is estimated as
 \begin{equation}\label{broad}
   \Delta^2=\frac{2\hbar^2 \omega_c}{\pi \tau}
 \end{equation}

Taking some typical numbers $B=5T$, $m=0.07m_e$, $\omega_c=2THz$ $\tau=4ps$, from Eq.(\ref{broad}) one finds that the  corresponding threshold frequency is of the order of $\sim 200GHz$. Therefore one can expect plateau-like behavior for frequencies below this threshold frequency.
 We think that the  plateau-like behavior observed in the range of $30-70GHz$ \cite{kuchar86,galchen87,dziom18}  is the manifestation of the above mentioned circumstances. 
 
 In the high frequency range $\omega\gg \Delta$  contribution of localized states close to complete filling of Landau level has an order of $\Delta^2/\omega^2\ll 1$ as it follows from Eq.(\ref{high}) and  is negligible. Therefore the filling of these states will not change the Hall conductivity and plateau-like behavior will be retained, see Fig.1. The plateau-like behaviour in the terahertz region observed in \cite{ikebe10,stier15,failla16} can be understood from the above mentioned point of view.
The ratio $\Delta^2/\omega^2$ in the above mentioned terahertz experiments $\omega\sim\omega_c$ could be less than $10^{-3}$ \cite{stier15}.








\end{document}